\documentclass[acmsmall,screen]{acmart} 
\usepackage{listings}
\usepackage{xcolor}
\usepackage{amsmath}

\usepackage{amssymb}
\usepackage{algorithm}
\usepackage{pifont}
\usepackage{algpseudocode}
\usepackage{multirow} 
\usepackage{graphicx} 
\usepackage{subfigure}
\usepackage[utf8]{inputenc}
\DeclareUnicodeCharacter{200B}{}

\newtheorem{theorem}{Theorem}
\newtheorem{definition}{Definition}

\AtBeginDocument{%
  }

\setcopyright{acmlicensed}
\copyrightyear{2024}
\acmYear{2024}
\acmDOI{XXXXXXX.XXXXXXX}

\acmISBN{978-1-4503-XXXX-X/18/06}

\begin{document}

\title{Petri Nets-based Methods on Automatically Detecting for Concurrency Bugs in Rust Programs}

\author{Kaiwen Zhang}
\email{zhangkw@tongji.edu.cn}
\author{Guanjun Liu}
\authornotemark[1]
\email{liuguanjun@tongji.edu.cn}
\affiliation{%
  \institution{Tongji university}
  \streetaddress{Caoan Road 2588}
  \state{Shanghai}
  \country{China}
  \postcode{180403}
}

\renewcommand{\shortauthors}{Zhang and Liu.}

\begin{abstract}
Rust's memory safety guarantees, notably ownership and lifetime systems, have driven its widespread adoption. Concurrency bugs still occur in Rust programs, and existing detection approaches exhibit significant limitations: static analyzers suffer from context insensitivity and high false positives, while dynamic methods incur prohibitive runtime costs due to exponential path exploration. This paper presents a Petri net-based method for efficient, precise detection of Rust concurrency bugs. The method rests on three pillars: (1) A syntax-preserving program-to-Petri-net transformation tailored for target bug classes; (2) Semantics-preserving state compression via context-aware slicing; (3) Bug detection through efficient Petri net reachability analysis.

 The core innovation is its rigorous, control-flow-driven modeling of Rust's ownership semantics and synchronization primitives within the Petri net structure, with data operations represented as token movements. Integrated pointer analysis automates alias identification during transformation. Experiments on standard Rust concurrency benchmarks demonstrate that our method outperforms the state-of-the-art methods LockBud and Miri that are both tools of detecting concurrency bugs of Rust programs. Compared to LockBud, our approach reduces false positives by 35.7\% and false negatives by 28.3\% , which is obtained through our precise flow-sensitive pointer analysis. Compared with Miri that is a dynamic analysis  tool, although Miri can obtain the same detection results, our method achieves 100× faster verification speed since our method takes a state reduce algorithm.
\end{abstract}

\ccsdesc[500]{Software and its engineering}
\ccsdesc[500]{Security and privacy~Trust frameworks}

\keywords{Concurrent bugs, Petri Net, Rust Language}

\maketitle

\section{Introduction}
Concurrent programming is essential for modern software performance but introduces non-deterministic defects-data races that evade traditional testing. Rust addresses this via its ownership system, where the borrow checker statically enforces memory safety and thread safety \cite{klabnik2019rust, matsakis2014rust}. This eliminates entire vulnerability classes without runtime overhead.

However, Rust’s guarantees remain bounded:Its type system cannot prevent high-level concurrency logic errors—deadlocks from cyclic lock dependencies, atomicity violations in different execution paths, or data races in unsafe blocks—which persist despite memory safety. Empirical evidence confirms that unsafe code blocks bypassing compiler checks proliferate in real-world Rust crates \cite{evans2020rust, zhang2024beyond}, reintroducing precisely the vulnerabilities Rust aims to eliminate.
Concurrency bug detection divided into static and dynamic analysis: ​Static methods​ identify bugs via predefined models such as lock graphs for deadlock detection or refinement types in theorem proving, yet suffer either high false positives due to oversimplified semantics (e.g., omitted control-flow paths) in pattern-matching methods, or poor scalability from labor-intensive modeling in proof assistants; ​Dynamic methods, while effective for observable runtime behaviors, incur coverage gaps that miss deep-seated flaws. To combine the advantages of both methods, we propose a ​Petri net-based method​ that semantic translation of program logic into net structures ensuring static rigor and enables exhaustive state exploration via reachability graphs achieving the effectiveness of dynamic verification. This resolves expressiveness limitations of ad-hoc models and state explosion in dynamic analysis, all without imposing manual modeling overhead.

However, existing Petri net-based methods \cite{petrinet2022, kindler2006petri, yao1997mapping, voron2008transforming, liao2013concurrency, rp, ding2009formal} for concurrency bug detection face three fundamental limitations:

1. Reliance on manual modeling of program logic into nets, incurring prohibitive overhead for industrial-scale projects. To address this problem, we propose an automated framework leveraging Rust’s Mid-Level IR (MIR):its control-flow graph inherently exposes execution paths, where data operations within basic blocks such as memory accesses directly map to token flows in Petri net places/transitions. By transforming each function into a self-contained subnet and composing the global net via call-graph topology, we establish a complete formal rule set including token consumption/regeneration for ownership transfers and lifetime constraints proven via equivalence to preserve Rust’s compile-time safety guarantees while enabling fully equivalent semantic transformations. 

2. The reliance on non-trivial extensions for data race detection via token-to-memory mappings drastically increases complexity. Using base Petri nets necessitates transforming each data operation individually, causing exponential net growth—a state explosion problem mirroring dynamic analysis: while dynamic methods suffer from combinatorial path explosion, Petri nets face marking-space explosion from model scaling. Our dual-strategy solution addresses this by:(1) ​Unsafe-centric transformation, selectively mapping only unsafe block data operations to places and extracting race-sensitive memory accesses via control-flow analysis, avoiding full-statement conversion; (2) ​Control-flow-driven state compression, merging synchronization-independent paths via program slicing and applying partial-order reduction to collapse symmetric states, reducing net size while preserving deadlock/atomicity violation detection completeness.  

3. Inaccurate place/transition mappings during automated translation, which manifest as erroneous concurrent access to shared resources, primarily stem from unverified aliasing relationships between variables accessing identical memory. Our solution introduces static alias-driven place binding through three key mechanisms: first, integration with Rust’s compiler to coalesce potentially aliased variables into identical places; second, generation of guard functions derived from Unsafe blocks, ensuring that multiple accesses to the same resource within a basic block trigger creation of only one identifying transition; and third, application of over-approximated analysis for may-alias scenarios like cross-function pointer aliasing. This conservative approach intentionally admits bounded false positives while guaranteeing the complete elimination of false negatives. 

The main contributions of this paper are as follows:
\begin{itemize}
	\item [1.] The first automated translation from Rust MIR to Petri nets via ownership-semantic-preserving token lifecycle rules, eliminating manual modeling while ensuring equivalence. 
	\item [2.] An extension-free concurrency bugs detection methodology using transition-labeling (read,write) in base Petri nets.
	\item [3.] A scalable verification engine integrating alias-analysis place binding and control-flow state compression.
\end{itemize}

The rest of this paper is organized as follows. Section \ref{background} gives an overview of Rust's concurrency model and the basic definition of Petri nets, giving a motivational example. Section \ref{framework} details the rules and algorithms for transforming Rust programs into Petri nets. Section \ref{algorithm} presents the algorithm for concurrent bug detection based on the state reachability graph of the Petri nets. The experimental results are presented in Section \ref{experimental} to demonstrate the effectiveness of the proposed method.  Section \ref{conclusion} concludes the paper.

\section{Related Work}
Despite Rust's compile-time memory safety guarantees, empirical studies reveal that unsafe code blocks which bypass compiler protections are pervasively adopted in real-world Rust libraries \cite{evans2020rust, mixed-study, xu2021memory, zhang2024beyond}, with significant portions of crates directly or transitively relying on unsafe operations even in prominent projects where low-level optimizations necessitate such implementations. Critically, these unsafe constructs reintroduce C/C++-style vulnerabilities including null pointer dereferences and memory access violations—precisely the risks Rust aims to eliminate. Static analyzers address memory safety issues but neglect concurrency vulnerabilities: Rudra \cite{bae2021rudra} detects unsafe API misuse through pattern matching, Mir-Checker \cite{li2021mirchecker} employs MIR-based symbolic execution to identify memory errors, and SafeDrop \cite{cui2023safedrop} utilizes path-sensitive data flow analysis for ownership bugs.

Specialized concurrency tools face inherent limitations: Miri's \cite{miri} dynamic analysis achieves precise data race detection via happens-before tracking but incurs prohibitive $O(n^2)$ thread interleaving overhead, restricting practical use to $\leq 4$ threads while failing to detect unexplored-path deadlocks. Conversely, Lockbud's \cite{qin2020understanding,qin2024understanding} static lock-graph approach predicts deadlocks but generates false positives/negatives due to inadequate path-sensitive alias analysis at synchronization points. Formal method adaptations struggle with Rust-specific challenges: theorem provers like Prusti \cite{astrauskas2022prusti}, RustBelt \cite{jung2017rustbelt} and Krust \cite{wang2018krust} require substantial code annotation overhead for refinement-type proofs, while model checkers suffer state explosion exacerbated by ownership semantics when naively translating to LLVM IR. 

Consequently, Rust concurrency verification persistently confronts dual challenges: (1) ​static analysis complexity, where pointer-analysis-dependent techniques require model restructuring for distinct bug patterns such as lock graphs needing redesign for condition variables, with existing models lacking native concurrency semantics support and thus forcing heuristic approximations in detection algorithms that degrade precision and induce high false positives\cite{ning2020stuck, 7582790, cai2021canary, cai2022peahen}; and (2) ​dynamic analysis limitations, as execution-dependent methods exhibit insufficient path coverage for long-running tasks\cite{dynamictrait, cai2021sound}, systematically omitting potential deadlocks in unexplored states.

\section{Background and Motivation}
\label{background}
\subsection{Rust Concurrency Model}
Rust's core safety mechanisms, particularly its ownership system and lifetime management, ensure memory safety at compile time. The ownership system operates on three fundamental rules: (1) Each value has a single owner; (2) Ownership moves when reassigned, invalidating the original variable; (3) Data access occurs through borrowing, with immutable references allowing shared access or mutable references for exclusive modification. For example:
\begin{verbatim}
let s1 = String::from("hello");
let s2 = s1;  // Ownership moves to s2, s1 becomes invalid
\end{verbatim}
Complementing ownership, lifetimes (\texttt{'a}) are compile-time annotations that manage reference validity, preventing dangling pointers by ensuring references don't outlive their data:
\begin{verbatim}
fn longest<'a>(x: &'a str, y: &'a str) -> &'a str {
    // Returned reference valid within 'a lifetime scope
}
\end{verbatim}

Rust's concurrency model builds upon these foundations through thread-level parallelism using \texttt{std::thread}, regulated by \texttt{Send} and \texttt{Sync} traits that govern cross-thread data transfer and sharing. The \texttt{Send} trait allows ownership transfer between threads, while \texttt{Sync} permits concurrent access through immutable references:
\begin{verbatim}
// Arc<T> implements Sync for shared immutable access
let data = Arc::new(Mutex::new(0));
\end{verbatim}
Despite these compile-time checks, concurrency bugs like deadlocks or atomicity violations under relaxed memory orderings (e.g., Relaxed) can still occur when synchronization primitives are misused.

Our analysis leverages Rust's Mid-level Intermediate Representation (MIR), which simplifies program syntax while preserving control flow and concurrency semantics. MIR decomposes programs into basic blocks representing discrete execution states, with concurrent program state modeled as their aggregation. This abstraction provides the foundation for detecting concurrency error patterns:
\begin{verbatim}
// MIR representation captures control flow and state transitions
bb0: {
    _1 = 42;                 // State assignment
    _2 = lock(_1);           // Concurrency operation
    ...
}
\end{verbatim}

\subsection{Petri Net}
A net is a 3-tuple $N=(P,T,F)$, where $P$ is a finite set of places, $T$ is a finite set of transitions, $F\subseteq (P\times T) \cup (T\times P)$ is a finite set of arcs, and $P\cap T=\emptyset$. A net can be viewed as a directed bipartite graph where places are represented by circles and transitions by rectangles. The pre-set and post-set of a node $x \in\{P \cup T\}$ are define as $^{\bullet}x=\{y\in P\cup T|(y,x)\in F\}$ and $x^{\bullet}=\{y\in P\cup T|(x,y)\in F\}$, respectively. In this paper, it is assumed that the weight of each directed arc is 1, that is, the firing of a transition consumes or generates only one token in a connected place. A marking of $N=(P,T,F)$ is a mapping $M:P\rightarrow \mathbb{N}$ where $M(p)$ is the number of tokens in place $p$. A net $N$ with an initial marking $M_{0}$ is called a Petri Net and denoted as $(N,M_{0})$. Transition $t$ is enabled at $M$ if $\forall p\in$$^{\bullet}t:M(p)>0$, denoted as $M[t\rangle$. A marking $M$ is a deadlock if any transition is disabled at $M$. Firing an enabled transition $t$ yields a new marking $M^{\prime}:M^{'}(p)=M(p)-1$ if $p\in^{\bullet}t\backslash t^{\bullet}$; $M^{\prime}(p)=M(p)+1$ if $p\in$$t^{\bullet}\backslash^{\bullet}t$; and otherwise $M^{\prime}(p)=M(p)$. This is denoted as $M[t\rangle M^{\prime}$. A marking $M_{k}$ is reachable from $M$ if $M_{k}=M$ or there exists a non-empty transition sequence $\sigma=t_{1}t_{2}\dots t_{k}$ such that $M[t_{1}\rangle M_{1}[t_{2}\rangle \dots M_{k-1}[t_{k}\rangle M_{k}$ (abbreviated as $M[\sigma \rangle M_{k}$). All markings reachable from $M_{0}$ in a Petri net is written as $\mathcal{M}$. 

The dynamic behavior of a Petri net is captured through its \textit{reachability graph}, which is a directed graph $RG = (V, E)$ defined as follows:
\begin{itemize}
    \item $V$ is the set of all reachable markings $\mathcal{M}$ starting from the initial marking $M_0$, i.e., $V = \mathcal{M}$.
    \item $E$ is the set of directed edges labeled with transitions $t \in T$, where $(M, t, M') \in E$ if and only if $M[t\rangle M'$.
\end{itemize}
The reachability graph provides a systematic way to explore all possible states (markings) of a Petri net and the transitions between them. Each node in $RG$ corresponds to a marking, and each directed edge represents the firing of a transition leading from one marking to another. If the reachability graph contains a cycle, it indicates the potential for repeated execution of certain transitions.

\subsection{A Motivating Example}
\begin{figure*}
    \centering
    \includegraphics[width=\textwidth]{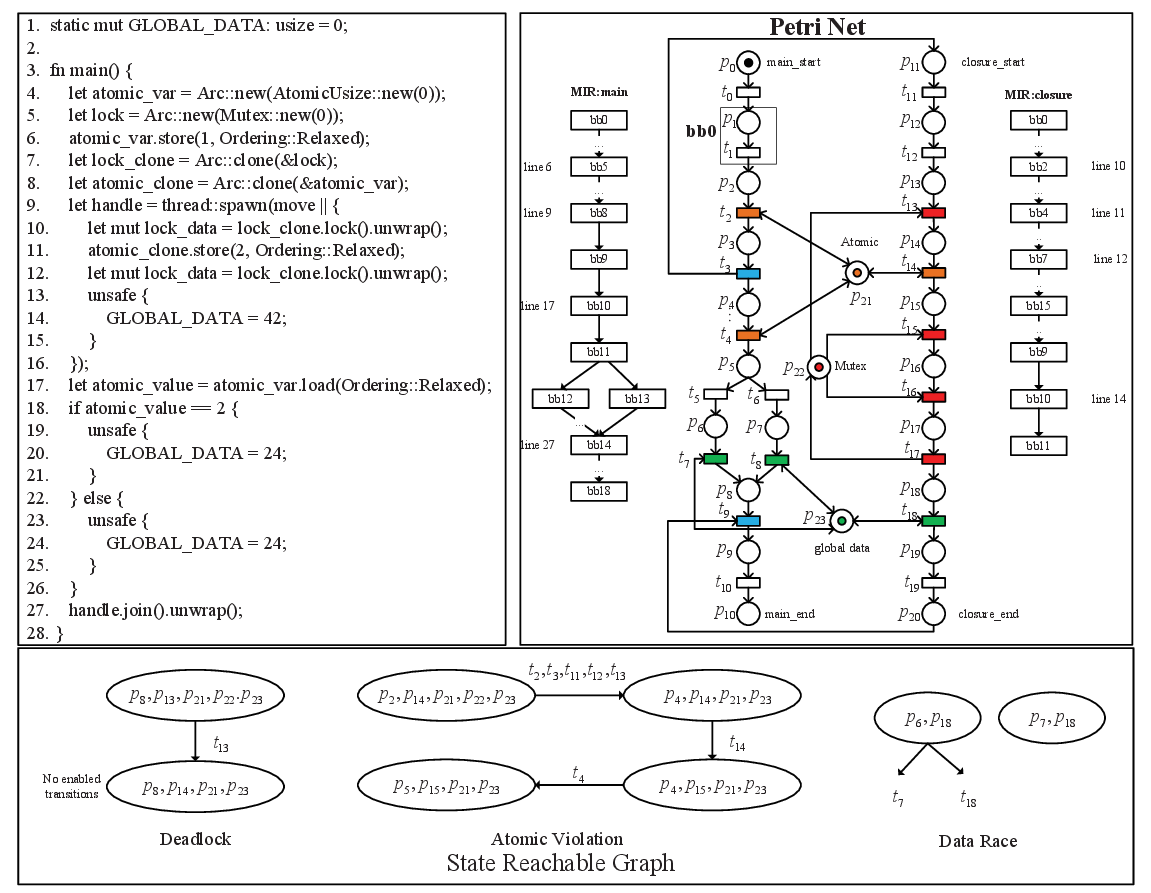}
    \caption{Illustration: Rust Code, MIR, Petri Net, and Bug Detection}
    \label{fig:motivate}
\end{figure*}

To detect concurrency bugs using Petri nets, the Rust program is first translated into a Petri net. This transformation is based on Rust's MIR, which simplifies the control flow and concurrency constructs. Using the Petri net, a reachability graph is generated to systematically explore all possible program states and identify bug patterns. The Fig. \ref{fig:motivate} provides a complete example, showing the source code, its corresponding MIR and Petri net representation, and the reachability graph with identified bug patterns such as deadlock, atomicity violation, and data race. Additionally, the MIR in Fig. \ref{fig:motivate} only includes the parts relevant to concurrency constructs; a complete MIR example can be found at \url{https://play.rust-lang.org/?version=stable&mode=debug&edition=2021}.

The Rust program in the upper left portion of Fig. \ref{fig:motivate} demonstrates multiple concurrency constructs, including atomic variables, mutexes, and unsafe global variable access. This program illustrates potential concurrency issues that are modeled in the corresponding MIR and Petri net representations. Key concurrency bugs in the program are identified as follows:
\paragraph{Deadlock:} A potential deadlock occurs in the spawned thread. Specifically, the second call to lock\_clone.lock() inside the closure attempts to acquire the mutex while the same thread already holds the lock. If another thread attempts to access the same mutex, a circular waiting condition may arise, leading to a deadlock.

\paragraph{Atomicity Violation:} The use of atomic\_var.store() and atomic\_var.load() with Ordering::Relaxed creates an atomicity violation. Relaxed memory ordering allows reordering of operations, which may cause inconsistent or unexpected behavior when threads read or write the atomic variable. This is particularly relevant in the main thread when the value of atomic\_var is checked.

\paragraph{Data Race:} Unsafe access to the global variable GLOBAL\_DATA introduces a data race. Since GLOBAL\_DATA is accessed and modified without synchronization in both the spawned thread and the main thread, multiple threads can concurrently read and write to it. This leads to undefined behavior and violates Rust's safety guarantees.
These issues are captured and analyzed in the Petri net representation and the corresponding reachability graph, as shown in the Fig. \ref{fig:motivate}. The reachability graph highlights states where these bugs manifest, such as the deadlock state and paths leading to atomicity violations or data races.

\section{Methodology}
\label{framework}

\begin{figure*}
    \centering
    \includegraphics[width=0.9\textwidth]{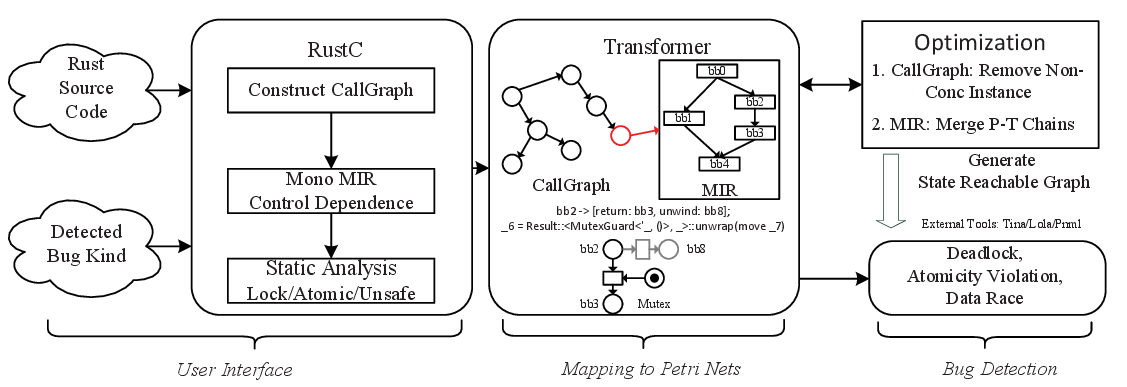}
    \caption{The framework of PPN}
    \label{fig:arch}
\end{figure*}
Rust concurrency errors mainly stem from uncontrolled contention on shared resources, including deadlocks, atomicity violations, and synchronization failures. Crucially, synchronization primitive operations occur in the terminator statement of the MIR basic block - the final instruction to which the control flow jumps, such as lock acquisition/release, atomic variable access. This allows us to focus on the terminator to capture synchronization points. However, for data race issues that may involve unsafe memory operations, we rely on explicitly marked Unsafe blocks in the Rust source code. When an Unsafe context is detected, the entire basic block is scanned to identify all memory access patterns and pointer operations.

To unify the semantics of control flow and resource interaction, we propose a program Petri net (PPN), a labeled extension of the basic Petri net. PPN solves two key problems through static analysis techniques: (1) parsing concurrency primitives (such as thread spawn, join, and resource scope) from the API call pattern of MIR; (2) identifying shared resource aliases through pointer analysis. This dual analysis ensures accurate modeling of inter-thread synchronization and shared memory dependencies.

The constructed PPN is streamlined to eliminate redundant states and transitions, improving analyzability while retaining the core behavioral semantics. The optimized net generates a reachability graph that supports systematic exploration of concurrent execution paths to detect bug modes such as deadlock, atomicity violation, and data race. Figure \ref{fig:arch} shows the entire process from MIR extraction to detect bugs.

\subsection{Program Representation}
We formally define a Rust program as a 6-tuple:
\begin{equation*}
    \mathbf{Prog} = (\mathcal{F}, G_{\text{call}}, \mathcal{CFG}, \Sigma, \mathcal{D}, \iota)
\end{equation*}
where each component is specified as follows:
\begin{itemize}
    \item[1.] Function Set $\mathcal{F}$: The set of monomorphized functions
    \begin{equation*}
        \mathcal{F} = \{ f_{1,\ldots ,n} \mid f \in \mathit{Decl} \} \cup \mathcal{C}
    \end{equation*}
    where $f_{1,\ldots,n}$ denotes generic functions instantiated with concrete types, $\mathcal{C}$ represents closure instances with captured environments.

    \item[2.] Call Graph $G_{\text{call}} = (\mathcal{F}, E_{\Lambda})$ with context-labeled edges:
    \begin{equation*}
        E_{\Lambda} \subseteq \mathcal{F} \times E_{t} \times \mathcal{F},\quad E_{t} = \{ \mathtt{spawn}, \mathtt{join}, \mathtt{scope}, \mathtt{normal\_call} \}
    \end{equation*}

    \item[3.] Control flow graph $\mathcal{CFG} = \{(bb_f, \delta_f, \mu_f)\}_{f\in\mathcal{F}}$ with $bb_f$ basic blocks, \\
        $\delta_f \subseteq bb_f \times \{Switch, Goto, \ldots, Return\} \times bb_f$, and $\mu_f: bb_f \rightarrow \Sigma^*$ satisfying $\forall b \in bb_f, \mu_f(b) = s_1;\dots;s_n \land s_n \in \mathit{Termination}$.

    \item[4.] Statement with Termination $\Sigma$ defined inductively:
    \begin{align*}
        \Sigma ::= &\ Call(f) \quad (f\in \mathcal{F}) \\
        &\ \mid Lock(m, \kappa) \mid UnLock(m) \quad (m\in \{Mutex, \ldots \},\kappa \in \{lock, read, \ldots \}) \\
        &\ \mid Atomic(ord) \quad (ord \in \{Relaxed, SeqCst, \ldots\}) \\
        &\ \mid Unsafe(\mathit{op}) \quad (op \in \{use, address,\ldots)
    \end{align*}

    \item[5.] Data Domain $\mathcal{D} = (\mathcal{M}, \mathcal{A}, \mathcal{U})$ where $\mathcal{M}$ is the set of synchronization resources, $\mathcal{A}$ is the set of atomic variables, $\mathcal{U}$ is the set of unsafe data.
    
    \item[6.] Entry Point $\iota = (\mathit{main}, \Gamma_0)$ with: $ \Gamma_0 = (\mathit{api}_0, \mathit{heap}_0, \mathit{panic\_handler})$.
\end{itemize}

It is important to note that the current version of Rust is still evolving, and many advanced features may not yet be fully supported by available tools, such as the latest language constructs or specific library functions. As a result, while our approach captures a large portion of Rust's core features, it deliberately omits niche syntax such as procedural macros or inline assembly to ensure the accuracy and feasibility of the conversion process through a simplified subset. This subset focuses exclusively on concurrency-critical elements, enabling rigorous conversion rules that maintain semantic correctness while reducing complexity. 

Within this methodology, we initially aggregate all monomorphized functions produced post-trait resolution to resolve generic types. Subsequently, we traverse this collection to construct an inter-procedural call graph. Given our focus on concurrency bug detection, function calls are categorically classified into concurrent calls and normal calls. The former abstracts thread control primitives including scope, spawn, and join as unified thread forking and synchronization actions. During call graph construction, we concurrently instrument basic blocks to monitor accesses to shared variables, specifically non-atomic mutable globals, alongside unsafe data operations. These observed patterns directly map to the targeted bug types. Furthermore, since Petri nets necessitate an initial state, we designate the program entry point, typically the main function, as the root node. Thread-spawning actions generate parallel transitions, while shared access points induce state conflicts. This methodology deliberately prioritizes verification tractability above full-language coverage, ensuring practical scalability, formal soundness in model translation, and adaptability to Rust's evolution through extensible annotation rules. Consequently, it enables comprehensive concurrent bug detection while minimizing ambiguities.

\subsection{Program Petri Net}
The PPN extends classical Petri nets by explicitly distinguishing between control flow and resource operations. This partitioning enables PPN to represent the dual nature of Rust programs, where control flow is governed by both sequential and concurrent transitions, and resource access is mediated through synchronization primitives, atomic variables, and unsafe operations. Notably, the incorporation of concurrency control transitions, which are associated with concurrency APIs, provides an important advantage: it allows us to identify functions that may execute concurrently. Its core innovation is that by binding control transitions to concurrent APIs, potential parallel execution domains are essentially identified, thereby automatically and accurately identifying concurrent function boundaries. In contrast, Lock-Graph assume that function calls are completely unordered interactions, which can easily produce high false positives due to path over-approximation; PPN uses the partial order modeling capability of Petri nets to explicitly characterize the dynamic interaction of thread forks, order constraints of synchronization events, and resource conflicts, fundamentally avoiding such false positives while maintaining verification scalability.

\begin{definition}
A PPN is a 5-tuple $(P, T, F, M, W)$ where:
\begin{itemize}
    \item[1.] $P = P_c \cup P_r$ represents the finite set of places: $P_c$ represent control places (program state nodes) while $P_r$ represent resource places (shared variables/resources).
    
    \item[2.] $T = T_c \cup T_o$ represents the finite set of transitions, specifically:
    $T_c = T_c^{\text{seq}} \cup T_c^{\text{conc}}$ where $T_c^{\text{seq}}$ represent sequential control operations (e.g., conditionals, loops) and $T_c^{\text{conc}}$ represent concurrency control operations;
    $T_o = T_o^{\text{sync}} \cup T_o^{\text{atom}} \cup T_o^{\text{unsafe}}$ where $T_o^{\text{sync}}$ represent synchronization primitives, $T_o^{\text{atom}}$ represent atomic operations, and $T_o^{\text{unsafe}}$ represent unsafe memory operations such as raw pointer dereference.
    
    \item[3.] $F \subseteq (P \times T) \cup (T \times P)$ represents the directed arcs, partitioned into control arcs $F_c \subseteq (P_c \times T_c) \cup (T_c \times P_c)$ and resource-operator arcs $F_{ro} \subseteq (P_r \times T_o) \cup (T_o \times P_r)$.
    
    \item[4.] $M: P \to \mathbb{N}$ represents the marking function with $M(p)$ denoting token count at place $p$, where $M_0$ specifies the initial marking.
    
    \item[5.] $W: F \to \mathbb{N}^+$ represents the weight function assigning positive integers to arcs.
\end{itemize}
\end{definition}

\subsection{Transformation from Rust Program to PPN}
\label{transformation}
The PPN-based modeling of Rust's concurrency APIs employs a three-layer translation framework leveraging control flow isomorphism ($\cong$) between CFGs and PPNs, preserving basic block semantics with minimal overhead:
\begin{itemize}
	\item[1.] ​CFG conversion​ bijectively maps each basic block to control place $p_b\in P_c$ representing program counter states during execution;
	\item[2.] Terminator-driven transition typing​ generates categorized transitions: branch terminators such as $switchInt$, synchronization primitives instantiate $T_o^{sync}$ with exclusion semantics, and unsafe scopes produce $T_o^{unsafe}$;
	\item[3.] Call graph-mediated inter-function linking​ transforms invocations into transitions: concurrent calls enable parallel token flow to $p_{entry_f}$ via $T_c^{conc}$, while sequential calls enforce token propagation through $p_{b^\prime}$.
\end{itemize}

Let $\mathcal{T}: \text{Prog} \to PPN$ be the transformation function. We formalize thread operations as:

\subsubsection{CFG Structural Mapping (Intra-function)}
\label{sssec:cfg-structure}

For each function $f \in \mathcal{F}$ with $\mathcal{CFG}_f = (bb_f, \delta_f, \mu_f)$:
\begin{align*}
\mathcal{T}(f) &= (P_f, T_f, F_f) \quad \text{where:} \\
P_f &= \{p_{\text{start}}^f, p_{\text{end}}^f\} \cup \{p_b \mid \forall b \in bb_f\} \\
T_f &= \{t_b \mid \forall b \in bb_f\} \cup \{t_{\delta} \mid \forall \delta \in \delta_f\} \\
F_f &= \{(p_{\text{start}}^f, t_{b_0})\} \cup \{(t_{\text{ret}}^{b_{\text{ret}}}, p_{\text{end}}^f)\} \quad \cup \bigcup_{b \in bb_f}  \{(p_b, t_b), (t_b, p_{b^\prime})\} 
\end{align*}

The transformation $\mathcal{T}$ converts a function $f$'s $\mathcal{CFG}_f = (bb_f, \delta_f, \mu_f)$ into a subnet of PPN  $(P_f, T_f, F_f)$ through element-wise mapping: the control places $P_f$ include the function entry $p_{\text{start}}^f$ and exit $p_{\text{end}}^f$ augmented by distinct places $p_b$ for each basic block $b \in bb_f$; the transitions $T_f$ consist of block activation transitions $t_b$ for every $b \in bb_f$ combined with terminator transitions $t_\delta$ encoding control flow decisions in $\delta_f$; the arcs $F_f$ establish connectivity initiating from $(p_{\text{start}}^f, t_{b_0})$ for the entry block, terminating at $(t_{\text{ret}}^{b_{\text{ret}}}, p_{\text{end}}^f)$ for return points, and chaining intra-block sequences via $\{(p_b, t_b), (t_b, p_{b^\prime})\}$ pairs that preserve CFG topology through rigorous place-transition correspondence, thus ensuring structural congruence between CFG paths and PPN firing sequences.

\subsubsection{Call Graph Mapping (Inter-function)}
\label{callgraph}
\begin{definition}[Core Transformation $\mathcal{T}_{\text{call}}$]
\[
\mathcal{T}_{\text{call}}(\delta_{\text{call}}, p_b, p_{b'}) = 
\begin{cases} 
t_{\tau} \in T_c^{\tau} \\
F_c \gets F_c \cup \Phi_{\tau}(p_b, p_{\text{start}}^f, p_{\text{end}}^f, p_{b'})
\end{cases}
\]
where $\tau = \tau(\delta_{\text{call}})$ and:
\begin{align*}
\Phi_{\text{Normal}} &= \{(p_b, t_{\tau}), (t_{\tau}, p_{\text{start}}^f), (p_{\text{end}}^f, t_{\text{ret}}), (t_{\text{ret}}, p_{b'})\} \\
\Phi_{\text{Spawn}} &= \{(p_b, t_{\tau}), (t_{\tau}, p_{\text{start}}^f), (p_{\text{end}}^f, \bot)\} \\
\Phi_{\text{Join}} &= \{(p_{\text{end}}^f, t_{\tau}), (t_{\tau}, p_{b'})\} \\
\Phi_{\text{Scope}} &= \left( \bigcup_{i=1}^n \Phi_{\text{Spawn}}(p_b, p_{\text{start}}^{f_i}, p_{\text{end}}^{f_i}, \bot) \right) \cup \{(t_{\text{sync}}, p_{b'})\} \cup \bigcup_{i=1}^n \{(p_{\text{end}}^{f_i}, t_{\text{sync}})\}
\end{align*}
\end{definition}

The transformation $\mathcal{T}_{\text{call}}$ formalizes the mapping through a type-driven methods $\tau(\delta_{\text{call}})$ that classifies call operations into four categories: Normal (synchronous function calls), Spawn (thread creation), Join (thread synchronization), and Scope (parallel scopes). For each type $\tau$, it generates a specialized transition $t_{\tau} \in T_c^{\tau}$ and extends the arc set $F_c$ via arc generator $\Phi_{\tau}$. Specifically: for Normal calls, $\Phi_{\text{Normal}}$ creates sequential path $p_b \to t_{\tau} \to p_{\text{start}}^f$ initiating callee execution followed by return path $p_{\text{end}}^f \to t_{\text{ret}} \to p_{b'}$ ensuring control returns to post-call block; for Spawn operations, $\Phi_{\text{Spawn}}$ establishes concurrent path $p_b \to t_{\tau} \to p_{\text{start}}^f$ launching parallel thread while marking callee exit with $\bot$ denoting detached termination; for Join synchronization, $\Phi_{\text{Join}}$ constructs blocking path $p_{\text{end}}^f \to t_{\tau} \to p_{b'}$ requiring thread completion token to proceed; for Scope constructs, $\Phi_{\text{Scope}}$ combines parallel spawn arcs $\bigcup_{i=1}^n \Phi_{\text{Spawn}}$ with barrier synchronization $\bigcup_{i=1}^n (p_{\text{end}}^{f_i} \to t_{\text{sync}}) \to p_{b'}$ enforcing all scoped threads complete before exit. This unified transformation preserves: 1) call-return semantics via $t_{\text{ret}}$ sequencing; 2) asynchronous parallelism via $\bot$-terminated paths; 3) blocking synchronization via $t_{\tau}$'s enabling condition; 4) collective synchronization via $t_{\text{sync}}$'s barrier semantics—all while maintaining structural congruence between CFG paths and PPN firing sequences.

\subsubsection{Lock and Unlock Operations}
\label{mutex}
$\mathcal{T}(\mathtt{mutex}) = 
\begin{cases} 
P_r \gets P_r \cup p_{mutex}, \ t \in T_o^{sync},\ M[p_{mutex}]\gets 1 \\
F_{ro} \gets F_{ro}  \cup 
\begin{cases} 
\{(p_{mutex}, t)\}, & \text{if } \mathtt{mutex} = \mathtt{lock} \\
\{(t, p_{mutex})\}, & \text{if } \mathtt{mutex} = \mathtt{drop}
\end{cases} 
\end{cases}$
\\ A mutex resource $p_{mutex}$ is created and initialized with a token ($M[p_{mutex}] = 1$). For a lock operation, the token is transferred from $p_{mutex}$ to the lock transition $t$, and for a drop operation the token is returned. This enforces exclusive access.

\subsubsection{Read-Write Lock Operations}
\label{rwlock}
$\mathcal{T}(\mathtt{rwlock}) = 
\begin{cases} 
P_r \gets P_r \cup p_{rwlock}, \ t \in T_o^{sync},\ M[p_{rwlock}]\gets n \\
F_{ro} \gets F_{ro} \cup 
\begin{cases} 
\{(p_{rwlock}, t, 1)\}, & \text{if } \mathtt{rwlock} = \mathtt{read} \\
\{(p_{rwlock}, t, n)\}, & \text{if } \mathtt{rwlock} = \mathtt{write} \\
\{(t, p_{rwlock}, n)\}, & \text{if } \mathtt{rwlock} = \mathtt{drop}
\end{cases} 
\end{cases}$
\\ 
Let $n\in N^+$ be the maximum number of concurrent readers allowed. The read-write lock resource $p_{rwlock}$ is initialized with $n$ tokens. A read operation consumes 1 token allowing up to $n$ concurrent readers, while a write operation consumes all $n$ tokens for exclusive access. The drop operation returns the tokens.

\subsubsection{Condition Variables Operations}
\label{condvar}
Condvar(lock) represents a condition variable and the lock resource associated with it. This transformation creates resource places for the lock ($p_{l}$), condition ($p_{cond}$), and waiting state ($p_{wait}$). The notify operation signals by transferring a token to $p_{cond}$, while the wait operation sequences transitions (via $t_{lo1}$, $t_{co}$, $t_{lo2}$) to model a thread waiting on the condition. This ensures proper synchronization between waiting and notification. \\
$\mathcal{T}(\mathtt{condvar(lock)}) = 
\begin{cases} 
P_r \gets P_r \cup \{p_{l}, p_{cond}, p_{wait}\}, \ t_{lo1}, t_{lo2} \in T_o^{sync},\
t_{co} \in T_o^{sync},\ M[p_{l}]\gets 1 \\
F_{ro} \gets F_{ro} \cup \{(t_{co}, p_{cond})\}, & \text{nofity} \\
F_{ro} \gets F_{ro} \cup \{(t_{lo1}, p_{l}),(t_{co}, p_{wait}),(p_{wait}, t_{lo2}),(p_{l}, p_{lo2},(p_{cond}, p_{lo2}))\}, & \text{wait}
\end{cases}$

\subsubsection{Bounded Channel Operation}
\label{channel}
$\mathcal{T}(\mathtt{channel}) = 
\begin{cases} 
P_r \gets P_r \cup p_{chan}, \ t \in T_o^{sync},\ M[p_{chan}]\gets n \\
F_{ro} \gets F_{ro} \cup 
\begin{cases} 
\{(t, p_{chan})\}, & \text{if } \mathtt{channel} = \mathtt{send} \\
\{(p_{chan}, t)\}, & \text{if } \mathtt{channel} = \mathtt{recv} 
\end{cases} 
\end{cases}$

\subsubsection{Atomic Variable Operations}
\label{atomic}
$\mathcal{T}(\mathtt{atomic}) = 
\begin{cases} 
P_r \gets P_r \cup p_{atomic}, \ t \in T_o^{atomic},\ M[p_{atomic}]\gets 1 \\
F_{ro} \gets F_{ro} \cup 
\{(t, p_{atomic}),(p_{atomic}, t)\}
\end{cases}$
\\ The atomic variable $p_{atomic}$ is created with a single token to guarantee atomic access. Both load and store operations are modeled by bidirectional transitions, ensuring that operations occur atomically and respect memory ordering.

\subsubsection{Unsafe Data Access}
\label{unsafe}
$\mathcal{T}(\mathtt{unsafe}) = 
\begin{cases} 
P_r \gets P_r \cup p_{unsafe}, \ t \in T_o^{unsafe},\ M[p_{unsafe}]\gets 1 \\
F_{ro} \gets F_{ro} \cup 
\{(t, p_{unsafe}),(p_{unsafe}, t)\}
\end{cases}$
\\ Unsafe data access is modeled using a dedicated resource $p_{unsafe}$ (initialized with a token). The bidirectional conversions allow read and write operations on insecure data; reads and writes do not change the state of the data.

\subsection{Algorithms for Petri Net Transformation }
\begin{algorithm}
\label{alg:sourceProg}
\caption{Source-to-Prog and Prog-to-PPN Conversion}
\begin{algorithmic}[1]
\Require Rust Source Code $RSC$
\Ensure Program Petri Net $PPN = (P, T, F, M_0, W)$

\State Phase 1: Build Prog model from source
\State \textbf{Initialize} $\mathcal{F} \gets \emptyset, G_{\text{call}} \gets (\emptyset, \emptyset), \mathcal{CFG} \gets \emptyset, \Sigma \gets \emptyset, \mathcal{D} \gets (\emptyset, \emptyset, \emptyset), \iota \gets \text{null}$
\State Compile $RSC$ to get monomorphized functions: $\mathcal{F} \gets \{\text{all } f_{\tau_1,\ldots,\tau_n}\}$

\ForAll{Function $f \in \mathcal{F}$} 
    \State Extract MIR as $\mathcal{CFG}_f = (bb_f, \delta_f, \mu_f)$
    \State $\mathcal{CFG} \gets \mathcal{CFG} \cup \{\mathcal{CFG}_f\}$
    
    \ForAll{Basic block $b \in bb_f$} 
        \ForAll{Statement $s \in \mu_f(b)$} 
            \State $\Sigma \gets \Sigma \cup \{s\}$
            \If{$s$ is resource operation}
                \If{$s = \mathtt{Lock}(m,\kappa)$ or $\mathtt{UnLock}(m)$}
                    \State $\mathcal{M} \gets \mathcal{M} \cup \{m, \kappa, f, s.location\}$ \Comment{Add to sync primitives}
                \ElsIf{$s = \mathtt{Atomic}(\ell, ord)$}
                    \State $\mathcal{A} \gets \mathcal{A} \cup \{a, ord, f, s.location\}$ \Comment{Add to atomic variables}
                \ElsIf{$s = \mathtt{Unsafe}(op)$}
                    \State $\mathcal{U} \gets \mathcal{U} \cup \{u, op, f, bb_f\}$ \Comment{Add to unsafe data}
                \EndIf
            \EndIf
        \EndFor
        
        \If{$s_{\text{term}}$ is call operation}
            \State Get callee $f_j$ and call type $\lambda$, $G_{\text{call}} \gets G_{\text{call}} \cup \{ (f, t, f_j) \}$
        \EndIf
    \EndFor
\EndFor

\State Run alias analysis on $\mathcal{D}$: $\mathcal{D} \gets \textsc{AliasAnalysis}(\mathcal{D})$
\State Set entry point: 
$\iota \gets \begin{cases} 
    (\text{main}, \Gamma_0) & \text{if binary program} \\
    (\text{API entries}, \Gamma_0) & \text{if library}
\end{cases}$

\State Construct $\mathbf{Prog} \gets (\mathcal{F}, G_{\text{call}}, \mathcal{CFG}, \Sigma, \mathcal{D}, \iota)$

\State Phase 2: Transform Prog to PPN
\State Apply Program-to-PPN Conversion (Algorithm \ref{alg:prog_to_ppnet}) on $\mathbf{Prog}$
\State \Return Resulting $PPN$
\end{algorithmic}
\end{algorithm}

The algorithm \ref{alg:sourceProg} constructs a simple program model Prog from Rust source code $RSC$ and converts it to a PPN. It comprises two phases: Phase 1 builds the Prog by compiling $RSC$ to obtain monomorphized functions $\mathcal{F}$, extracting the MIR into CFG for each function $f \in \mathcal{F}$, and iterating over all basic blocks and statements to collect program elements such as synchronization primitives $\mathcal{M}$, atomic variables $\mathcal{A}$ , and unsafe data $\mathcal{U}$ while handling resource operations including locks, atomic accesses, and unsafe statements; it then performs alias analysis on dependency data $\mathcal{D}$ and sets the entry point $\iota$ based on whether the program is a binary or a library. Phase 2 invokes the Algorithm \ref{alg:prog_to_ppnet} to transform the constructed Prog model into the final PPN, ensuring the conversion preserves the program's concurrent behavior semantics for formal verification.

\begin{algorithm}
\caption{Program-to-PPN Conversion}
\label{alg:prog_to_ppnet}
\begin{algorithmic}[1]
\Require Program $\mathbf{Prog} = (\mathcal{F}, G_{\text{call}}, \mathcal{CFG}, \Sigma, \mathcal{D}, \iota)$
\Ensure PPN $PPN = (P, T, F, M_0, W)$
\State \textbf{Initialize} $P \gets \emptyset, T \gets \emptyset, F \gets \emptyset, M_0 \gets \emptyset, W \gets \emptyset$

\ForAll{Function $f \in \mathcal{F}$} 
    \State Add function entry place and return place: $P \gets P \cup \{p_{entry}^f, p_{exit}^f\}$
    \State Add start/return transitions: $T \gets T \cup \{t_{start}^f, t_{return}^f\}$
    \State Add arcs: $F \gets F \cup \{(p_{entry}^f, t_{start}^f), (t_{return}^f, p_{exit}^f)\}$
 
    \ForAll{Basic block $b \in bb_f$} 
        \State Add control places and transitions $p_b:P \gets P \cup \{p_b\}$, $t_{b}:T \gets T \cup \{t_{b}\}$
        \State Add arcs: $F \gets F \cup \{(p_{b}, t_{b})\}$
    \EndFor
    \State Add arc: $F \gets F \cup \{(t_{start}^f, p_{b_0})\}$
  
    \ForAll{Basic block $b \in bb_f$} 
        \If{b is \texttt{unsafe}-marked}
            \ForAll{Statement $s \in \mu_f(b)$} 
                \State Map to resource operation using Rules \ref{unsafe}
            \EndFor
        \ElsIf
            \State Let $s_{\text{term}} = \text{terminator of } b$
            \If{$s_{\text{term}}$ is a Call operation}
                \State Add call arcs via $G_{\text{call}}$ edges (Rule \ref{callgraph})
            \ElsIf{$s_{\text{term}}$ is a Sync primitive}
                \State Map via Rules \ref{mutex}, \ref{rwlock}, \ref{condvar}, \ref{channel}
            \ElsIf{$s_{\text{term}}$ is an Atomic operation}
                \State Map via Rule \ref{atomic}
            \EndIf
        \EndIf
    \EndFor
\EndFor

\State Set initial marking for entry points: $\forall p \in API_{\text{init}}, M_0(p) \gets 1$
\State \Return $PPN = (P, T, F, M_0, W)$
\end{algorithmic}
\end{algorithm}


This algorithm \ref{alg:prog_to_ppnet} transforms the Prog into a PPN by iterating over all functions $f \in \mathcal{F}$ to add function entry and exit places ($p_{\text{entry}}^f, p_{\text{exit}}^f$) along with corresponding transitions ($t_{\text{start}}^f, t_{\text{return}}^f$); it then processes each basic block $b$ to introduce control places ($p_b$) and transitions ($t_b$), connecting them via arcs based on the control flow. Subsequent iterations handle the terminator statements $s_{\text{term}}$: if a block is marked unsafe, unsafe statements are mapped using predefined rules; otherwise, call operations are resolved through the call graph $G_{\text{call}}$, synchronization primitives map to mutexes, read-write locks, condition variables, or channels, and atomic operations utilize specific mapping rules. Finally, the algorithm sets the initial marking $M_0$ on entry places to represent initial program states, generating a PPN that models data dependencies and concurrency patterns for static analysis of program behaviors.

The complexity of the algorithm depends on the size of the CFG, the function call graph, and the number of resources and operations. Let $|B|$ be the number of basic blocks and $|E|$ be the number of edges in the CFG, $|F|$ the number of functions, and $|R|$ the number of resources, each with $|O|$ operations. Constructing the control flow has a complexity of $O(|B| + |E|)$, and resource operations require $O(|R| \cdot |O|)$. Therefore, the overall complexity of the algorithm is $O((|B| + |E|)\cdot |F| + |R| \cdot |O|)$. This complexity ensures that the algorithm is efficient and scalable for typical program sizes, making it suitable for practical analysis of concurrent systems.

\section{Detect General Concurrency Bugs}
\label{algorithm}
\subsection{Deadlock Detection}
Deadlocks represent a critical class of concurrency bugs where threads are permanently unable to proceed due to cyclic resource dependencies. Petri nets utilize transition firing rules to model state changes, dynamically capturing all possible behaviors of a program. However, previous research on deadlocks in Petri nets has focused predominantly on global deadlocks, which are defined as the termination of the entire program. A deadlock state is defined as a state where no transitions are enabled under non-terminal markings. For the detection of local deadlocks, more precise transition definitions and complex logical formulas are required. Local deadlocks in the reachability graph are often represented as cycles, where certain shared resources never receive tokens. We classify this type of deadlock as a "deadlock class" in which some locks or resources remain perpetually unavailable. To formalize, if there exists a cycle in the reachability graph where, within the cycle, a resource remains at zero tokens indefinitely, we define this as a local deadlock. In our approach, we combine the detection algorithms for these two types of deadlocks. If the program reaches a terminal mark, the global deadlock detection method is applied. If the program does not reach a terminal marking, the local deadlock detection method is used.

We consider two types of deadlock pattern within our methods:
\begin{itemize}
  \item [1.] A \emph{global deadlock} is formally defined as a marking $M$ satisfying two necessary and sufficient conditions: (1) no transitions are enabled ($\text{Enabled}(M) = \emptyset$), indicating the impossibility of any further state evolution, and (2) $M$ does not constitute a valid termination state $M_{\text{final}}$ – where termination is explicitly characterized by the concurrent fulfillment of two requirements: (i) complete termination of the \texttt{main} function's execution, and (ii) restoration of all resource places to their initial token capacities ($\forall p \in P_{r}, M(p) = \text{Init}(p)$); consequently, the formal deadlock criterion is expressed as $\text{Enabled}(M) = \emptyset \land M \neq M_{\text{final}}$.
  \item [2.] A \emph{local deadlock} denotes a partial system failure where one or more threads become indefinitely blocked due to resource contention while other threads continue normal execution—a condition undetectable by global deadlock methods. This manifests in three distinct scenarios: (1) program termination occurs without recognizing perpetually blocked threads (functionally resembling undetected global deadlock); (2) active threads suspend execution waiting for blocked threads to release resources, ultimately triggering global deadlock; or (3) the system enters cyclic behavior whereby it periodically revisits identical states in the reachability graph. Crucially, local deadlocks are formally identified in cyclic scenarios when at least one lock-related resource place $p_r$ exhibits persistent token starvation ($\forall M_i \in \text{cycle },\ M_i(p_r) = 0$) and never regains its initial token capacity, indicating irrevocable resource seizure preventing thread progression through the cycle.
\end{itemize}
This algorithm \ref{alg:deadlock_detection} employs a tiered detection architecture, prioritizing computational resources for deadlock patterns while ensuring rare but dangerous partial deadlocks are captured. Model reduction during preprocessing downgrades state space complexity, solving the state explosion problem. Innovatively integrates three deadlock scenarios into a unified framework: Scenario 1 through post-termination thread resource auditing, Scenario 2 via standard global detection, and Scenario 3 through optimized cycle analysis using SCC decomposition.

\begin{algorithm}
\caption{Deadlock Detection in PPN}
\label{alg:deadlock_detection}
\begin{algorithmic}[1]
\Require Reachability graph $RG = (V,E)$ with $|M| = n$ markings, 
         \par \hspace{\algorithmicindent} Resource places $P_r$ with initial capacities $\text{Init}(p_r)$, final marking classes: $M_{final}$
\Ensure Set $DM$: global deadlocks + local deadlock cycles
\State $DM \gets \emptyset$
\State Apply model reduction (Sec.~\ref{reducepn}) to compress $\mathcal{G}$

\State \textsc{Phase 1: Global \& Scenario 1 Detection}
\ForAll{$M \in RG$} 
    \If{$\text{Enabled}(M) = \emptyset$} 
        \If{$M \in \mathcal{M}_{\text{final}}$}          
            \If{$\exists p_r^t \in \{^{\bullet}(t_i),(t_i)^{\bullet}\}: M(p_r^t) \neq \text{Init}(p_r^t)$} 
                \State $DM \gets DM \cup \{\langle M, t_i\rangle\}$ \Comment{Scenario 1}
            \EndIf
        \Else 
            \State $DM \gets DM \cup \{M\}$ \Comment{Global deadlock (Covers Scenario 2)}
        \EndIf
    \EndIf
\EndFor

\State \textsc{Phase 2: Cycle Analysis for Scenario 3}
\State Compute SCCs using Tarjan's algorithm 
\ForAll{cycle $C\subset nontrivial SCC\in RG$}
        \ForAll{$p_r \in P_r$ with $\text{usedInCycle}(C, p_r)$} 
            \If{$\forall M_i \in C, M_i(p_r) = 0$}
                \If{$\nexists t \in T_{unlock} \text{ enabled in } C$ \textbf{and not} \textsc{BenignCycle}(C)} 
                    \State $DM \gets \mathcal{D} \cup \{C\}$ \Comment{Scenario 3}
                \EndIf
            \EndIf
    \EndFor
\EndFor
\State \Return $DM$ 
\end{algorithmic}
\end{algorithm}

\subsection{Data Race Detection}
\label{datarace}
Data races in Rust manifest themselves primarily through two mechanisms: (1) explicit unsafe operations on global static data through raw pointers; and (2) logical flaws in manually implemented Sync traits that violate thread safety guarantees. Both patterns bypass borrow rule checks via the Unsafe annotation, formally defined in this study as concurrent unsynchronized accesses to a shared memory location where at least one access is a write. Within the PPN, data races are detected by analyzing the markings in the state reachability graph, where multiple enabled transformations concurrently access the same unsafe data location $p_{unsafe} \in P_{r}$ with either a read-write conflict or a write-write conflict. Crucially, our model distinguishes benign shared reads from dangerous write conflicts through transformation metadata annotations.

\begin{definition}[Data Race]
Given a RG of a PPN $N$, a marking $M$ is said to exhibit a data race on an unsafe data place $p_{unsafe} \in P_r$ if there exist two distinct transitions $t_1, t_2 \in T$ such that:
\begin{itemize}
    \item Both $t_1$ and $t_2$ are enabled in $M$: $t_1, t_2 \in \text{Enabled}(M)$.
    \item Both transitions access the same unsafe data place $p_{\text{unsafe}}$.
    \item At least one of the transitions is a write operation.
\end{itemize}
\end{definition}

\begin{algorithm}
\caption{Data Race Detection in PPN}
\label{alg:data_race_detection}
\begin{algorithmic}[1]
\Require Reachability Graph $RG = (V, E)$, Unsafe Data Places $P_{\text{unsafe}}$
\Ensure Data Race Reports $\mathcal{R}$
\State $\mathcal{R} \gets \emptyset$
\ForAll{$M \in RG$}
    \State $\text{Enabled}(M) \gets \{t \in T \mid t \text{ is enabled in } M\}$
    \ForAll{$p_{unsafe} \in P_{unsafe}$}
        \State $\text{ConflictSet} \gets \{(t_1, t_2) \mid t_1, t_2 \in \text{Enabled}(M), \, t_1 \neq t_2, \, p_{\text{unsafe}} \in ^{\bullet}t_1 \cap ^{\bullet}t_2,$ 
        \Statex \hspace{6.5cm} $\text{and at least one of } t_1, t_2 \text{ is a write}\}$
        \ForAll{$(t_1, t_2) \in \text{ConflictSet}$}
            \State $\text{Loc}_1 \gets \text{Metadata}(t_1)$, $\text{Loc}_2 \gets \text{Metadata}(t_2)$ 
            \If{$\text{Loc}_1 = \text{Loc}_2$} 
                \State \textbf{continue} \Comment{Filter same-location conflicts}
            \EndIf
            \State $\mathcal{R} \gets \mathcal{R} \cup \{(p_{\text{unsafe}}, M, t_1, t_2)\}$ 
        \EndFor
    \EndFor
\EndFor
\State \Return $\mathcal{R}$
\end{algorithmic}
\end{algorithm}

This algorithm \ref{alg:data_race_detection} identifies data races in a PPN using the reachability graph. For each marking $M$, it computes the set of enabled transitions $\text{Enabled}(M)$ and iterates over all unsafe data places $p_{unsafe}$. A conflict set is constructed by identifying pairs of enabled transitions $(t_1, t_2)$ that operate on the same unsafe data place and where at least one transition performs a write operation. To enhance the precision and clarity of bug reports, operations originating from the same unsafe block or associated with the same code location are grouped and aggregated into a single bug report. This aggregation reduces redundant reporting of data races caused by the same unsafe access pattern, making the detection process more interpretable and actionable for developers.

\subsection{Atomicity Violation Detection}
\label{atomic}
Atomicity violations occur when the intended atomic behavior of operations on shared atomic variables is violated. In multithreaded programs, these violations arise when the execution of operations on shared variables is interrupted or interleaved with operations from other threads, leading to inconsistent states. In this work, we adopt the general atomicity violation pattern from \cite{cai2021sound}.

\begin{definition}[Atomicity Violation: Unsynchronized Load-Store]
Given a PPN and an atomic variable place $p_{\text{atomic}} \in P_r$, an atomicity violation occurs when:
\begin{itemize}
    \item There exists a relaxed load operation $t_{load}$ on $p_{atomic}$.
    \item Two distinct relaxed store operations $t_{store_1}$ and $t_{store_2}$ occur on the same atomic variable $p_{atomic}$.
    \item The relaxed load operation $t_{load}$ is preceded by at least two distinct relaxed load operations $t_{store_1}$ and $t_{store_2}$.
    \item The relaxed load operation $t_{load}$ is followed by a branching operation, such as a conditional if or switch, or implicit control flow due to function calls.
\end{itemize}
\end{definition}

\begin{algorithm}
\caption{Atomicity Violation Detection in PPN}
\label{alg:atomic_rg}
\begin{algorithmic}[1]
\Require Reachability Graph $RG=(V,E)$, $P_{\text{atomic}}$, $T_{\text{branch}}$
\Ensure Violation set $\mathcal{V}$
\State $\mathcal{V} \gets \emptyset$
\State Build transition-state map $\mathcal{M}_T(s) \gets \{t \mid (v', v, t) \in E\}$ 
\ForAll{$v_{\text{load}} \in V$ where $\exists t_{\text{load}} \in T_{\text{relaxed-load}}$ fired at $v_{\text{load}}$}
    \If{$\exists v_{\text{next}}$: $(v_{\text{load}}, v_{\text{next}}, v_{\text{branch}}) \in E$} \Comment{Branch after load}
        \State $\text{Stores} \gets \emptyset$
        \State $\text{Visited} \gets \emptyset$
        \State $\text{Backtrack}(s_{\text{load}}, p_{\text{atomic}}, \text{Stores}, \text{Visited})$
        \If{$|\{t_s \in \text{Stores} \mid t_s \text{ is store}\}| \geq 2$}
            \State $\mathcal{V} \gets \mathcal{V} \cup \{(p_{\text{atomic}}, s_{\text{load}}, \text{Stores})\}$
        \EndIf
    \EndIf
\EndFor
\State \Return $\mathcal{V}$

\Procedure{Backtrack}{$s$, $p_a$, $\text{Stores}$, $\text{Visited}$}
\If{$s \in \text{Visited}$} \textbf{return} \EndIf
\State $\text{Visited} \gets \text{Visited} \cup \{s\}$
\ForAll{$t \in \mathcal{M}_T(s)$} \Comment{Incoming transitions}
    \If{$t$ accesses $p_a$ \textbf{and} $t \in T_{\text{relaxed-store}}$}
        \State $\text{Stores} \gets \text{Stores} \cup \{t\}$
    \ElsIf{$t$ is synchronization} \textbf{continue} \Comment{Stop at sync}
    \EndIf
    \State $\text{Backtrack}(\text{source}(t), p_a, \text{Stores}, \text{Visited})$ 
\EndFor
\EndProcedure
\end{algorithmic}
\end{algorithm}


This algorithm \ref{alg:atomic_rg} detects atomicity violations exclusively via Petri net reachability graph, motivated by three pivotal insights: atomicity violations manifest as specific path patterns ($\leq 2$ relaxed stores between a relaxed load and its control-dependent branch); synchronization operations break these patterns and must terminate backtracking; and RG's sparsity (average in-degree $d \ll |\mathcal{M}|$) enables efficient traversal. Thus, it constructs state-input transition map $\mathcal{M}_T$ for $O(1)$ predecessor lookup, recursively backtracks from each relaxed-load state transition $s_{\text{load}}$ to collect relaxed stores $t_s$ accessing the same atomic place $p_a$, reporting violations when: 1) subsequent branch transition $t_{\text{branch}}$ exists; 2) $\leq 2$ distinct $t_s$ are found; 3) no synchronization intervenes. 

\subsection{Complexity Analysis of Detection Algorithms}
The complexity of detecting deadlocks, data races, and atomicity violations depends on the structure of the RG and the PPN. Let $|E|$ the number of edges in the RG, $|P|$ the number of places, $|T|$ the number of transitions, and $|F|$ the number of arcs in the PPN. The complexity for each detection algorithm is analyzed as follows:
\begin{itemize}
    \item [1.] \textbf{Global Deadlock Detection:}  
    The global deadlock detection algorithm iterates over all reachable states in the RG. For each state, the algorithm checks whether the set of enabled transitions is empty. This requires a traversal of all $|\mathcal{M}|$ states in the RG, with each state requiring $O(|T|)$ time to determine enabled transitions. Thus, the overall complexity of global deadlock detection is:
    \[
    O(|\mathcal{M}| \cdot |T|)
    \]
    The space complexity is $O(|\mathcal{M}|)$, as all markings in the RG must be stored.

    \item [2.] \textbf{Cycle Deadlock Detection:}  
    Detecting cycle deadlocks involves identifying all cycles in the RG and analyzing the transitions within each cycle. For a given cycle, the algorithm checks if there exists a resource place that remains without tokens for the entire cycle. Detecting cycles in the RG requires $O(|E| + |\mathcal{M}|)$ time using Tarjan's strongly connected components (SCC) algorithm. For each cycle, transitions are checked to determine if resource places are perpetually empty. In the worst case, every cycle involves all transitions, resulting in a complexity of $O(|\text{cycles}| \cdot |T|)$. Therefore, the total complexity of cycle deadlock detection is:
    \[
    O((|\mathcal{E}| + |\mathcal{M}|) + |\text{cycles}| \cdot |T|)
    \]
    Space complexity is dominated by storing cycles and their associated transitions, which is proportional to the number of cycles.

    \item [3.] \textbf{Data Race Detection:}  
    The data race detection algorithm analyzes all states in the RG. For each state, the enabled transitions are identified, and all pairs of enabled transitions are compared to detect conflicting operations on shared `unsafe` data places. This requires $O(|T|^2)$ comparisons per state, leading to a total complexity of:
    \[
    O(|\mathcal{M}| \cdot |T|^2)
    \]
    Additional filtering to group operations by unsafe blocks or code locations involves metadata checks, which have negligible overhead compared to the pairwise transition comparisons. The space complexity is $O(|\mathcal{M}|)$ for storing state information.

    \item [4.] Atomicity Violation Detection: 
    Detecting atomicity violations involves analyzing transitions labeled as Relaxed Load or Relaxed Store. For each Relaxed Load, the algorithm performs a backward traversal of the RG to identify all preceding transitions affecting the same resource place. Identifying Relaxed Load transitions requires $O(|T|)$ time, and for each such transition, a backward traversal over the RG has a complexity of $O(|E|)$. Thus, the total complexity is:
    \[
    O(|T| \cdot |E|)
    \]
    The space complexity is $O(|\mathcal{M}|)$, as it depends on the storage of visited states during backward traversal.
\end{itemize}

\subsection{IMPROVE PERFORMANCE}
\label{reducepn}
To improve the efficiency of detection algorithms, such as those for state reachability and concurrency bug detection, we simplify Petri nets by reducing the number of places and transitions while preserving their behavioral properties. This is crucial to mitigate the state space explosion problem, which often hampers analysis in complex systems. In Algorithms \ref{alg:pn_reduction}, we present three key reduction methods:

Linear Sequence Simplification: For branchless place-transition alternating paths (denoted as $p_i \to t_1 \to p_2 \to \dots \to p_k$), compression is applied when path length exceeds threshold $L_{\min}$ (typically 3 nodes) and every internal transition has exactly one input and output arc ($indeg(t_i) = outdeg(t_i) = 1$). The reduction identifies endpoints $p_i$ and $p_k$ via connectivity analysis, creates new transition $t_{new}$ representing the sequence, updates flow relation to $F \leftarrow (F \setminus \{(p_i,t_1),(t_1,p_2),\dots,(t_n,p_k)\}) \cup \{(p_i,t_{new}),(t_{new},p_k)\}$ with weight aggregation, and removes all intermediate nodes $\{p_2,\dots,p_{k-1}, t_1,\dots,t_n\}$, preserving firing sequence equivalence while reducing state space dimensionality.

Concurrency-Irrelevant Call Chain Removal: This method eliminates nodes exclusively belonging to sensitive paths (those containing $\geq 1$ transition from $T_c^{\text{conc}}$ or place adjacent to $P_r$) while preserving nodes shared with non-sensitive paths. All paths from $p_{\text{start}}$ to $p_{\text{end}}$ are classified into sensitive ($\pi_S$) and non-sensitive ($\pi_N$) sets. For each node $v$ in sensitive paths $\pi \in \pi_S$, if $v \notin \bigcup_{\pi_N} \{v | v \in \pi_N\}$ and $v \neq p_{\text{start}}$, it is marked for deletion. Marked nodes and adjacent arcs are removed, with connectivity maintained through $p_{\text{start}}$'s external links, ensuring only path-exclusive sensitive segments are eliminated without affecting shared components.

Non-Concurrent Loop Optimization: For each loop $l$ within function-derived subnets $\mathcal{F}$, when its transitions $T_L$ contain no concurrency operations ($T_L \cap T_c^{conc} = \emptyset$) and places $P_L$ contain no resource places ($P_L \cap P_r = \emptyset$), the cycle is broken by removing exactly one back-edge $(t_{last}, p_{first})$ where $t_{last}$ is the final transition before looping and $p_{first}$ is the cycle entry point, preserving all external connections to loop nodes while eliminating cyclic behavior through minimal structural modification.

\begin{algorithm}
\caption{Petri Net Reduction Algorithm}
\label{alg:pn_reduction}
\begin{algorithmic}[1]
\Require Program Petri net $PPN = (P, T, F, W, M_0)$, resource places $P_r$, concurrent transitions $T_c^{conc}\in T$, entry point:$\{p_{start},p_{end}\}$
\Ensure $PPN' = (P', T', F', W', M_0')$ 
\Procedure{ReducePetriNet}{}
    \State \textproc{CompressLinearSequences}($PPN$)
    \State \textproc{PruneSensitiveChains}($PPN, p_{start}, p_{end}, T_c^{conc}, P_r$)
    \State \textproc{ReduceLocalLoops}($PPN, T_c^{conc}, P_r$)
    \State \Return $PPN$
\EndProcedure

\Function{CompressLinearSequences}{$PPN$}
    \ForAll{maximal linear path $\pi = [p_s \rightarrow t_1 \rightarrow \cdots \rightarrow p_k]$}
        \If{$|\pi| \geq 3$ \textbf{and} $\forall t_i \in \pi\colon \textit{indeg}=\textit{outdeg}(t_i) = 1$}
            \State Replace $\pi$ with $p_s \rightarrow t_{\mathit{new}} \rightarrow p_k$
            \State Remove intermediate nodes $\{p_2,\ldots,p_{k-1}, t_1,\ldots,t_{|\pi|/2}\}$
        \EndIf
    \EndFor
\EndFunction

\Function{PruneSensitiveChains}{$PPN, p_{start}, p_{end}, T_c^{conc}, P_r$}
    \State Sensitive paths:$\pi_S \gets \emptyset$, non-sensitive paths:$\pi_N \gets \emptyset$
    \For{place-transition paths:$\pi_{pt} \in [p_{start} \rightarrow t_1 \rightarrow p_1 \rightarrow \cdots \rightarrow p_{end}]$}
        \If{$\exists t \in \pi_{pt}: (t \in T_c^{conc} \vee p=\{^{\bullet}t\cup t^{\bullet}\} \in P_r)$}
            \State $\pi_S \gets \pi_S \cup \{\pi_{pt}\}$
        \Else
            \State $\pi_N \gets \pi_N \cup \{\pi_{pt}\}$
        \EndIf
    \EndFor

    \For{$\pi \in \pi_S$}
        \For{each node $v \in \pi$}
            \If{$v \notin \bigcup_{\pi_{pt} \in \pi_N} \{v | v \in \pi_{pt}\}$ \textbf{and} $v \neq p_{start}$}
                \State Mark $v$ for deletion: $V_{del} \gets V_{del} \cup \{v\}$
            \Else
                \State \textbf{break} 
            \EndIf
        \EndFor
    \EndFor
    
    \State Remove all $v \in V_{\text{del}}$ and adjacent arcs
\EndFunction

\Function{ReduceLocalLoops}{$PPN, T_c^{conc}, P_r$}
    \For{each function subnet $f \in \mathcal{F}$}
        \For{each loop $l \in L$}
            \State Extract $P_L, T_L$ within $l$
            \If{$T_L \cap T_c^{conc} = \emptyset$ \textbf{and} $P_L \cap P_r = \emptyset$}
                \State Break loop by removing $(t_{\textit{last}}, p_{\textit{first}}) \in F$
            \EndIf
        \EndFor
    \EndFor
\EndFunction

\end{algorithmic}
\end{algorithm}

\section{Experimental Evaluation}
\label{experimental}
\subsection{Experiment Setup and Benchmarks}
To rigorously assess our Petri net-based framework against state-of-the-art tools (LockBud and Miri), we designed three benchmark sets spanning controlled tests and real-world Rust projects: 
\textbf{L-Bench} (LockBud's test suite), \textbf{M-Bench} (Miri's concurrency tests), and \textbf{P-Bench} (our curated projects including \textit{Rayon-demo} (24k~LoC) and \textit{Openethereum} (128k~LoC)). These benchmarks measure detection capabilities for deadlocks, data races, and atomicity violations through metrics including real bugs (RB), false positives/negatives (FP/FN), and time efficiency. Our results demonstrate that integrating structural analysis—enhancing classical Petri nets with constraint arcs for memory ordering and hierarchical subnets for contextual lock modeling—with state reachability exploration—using partial-order reduction to prune redundant interleavings—overcomes inherent Petri net limitations in detecting local deadlocks and atomicity violations. For deadlocks, traditional approaches ignored dynamic contexts, causing LockBud's false positives in lock-closure (FP=1), while our hybrid method achieved 100\% in inter (12/12~RBs detected) versus Miri's 91.7\% false negative rate. For atomicity violations, LockBud misflagged safe SeqCst operations in atomic-SeqCst (FP=4) due to unmodeled memory semantics, whereas our constraint arcs embedded Rust's memory model, enabling precise detection of all true violations in atomic-violation (4/4~RBs). State exploration further exposed Miri's coverage gaps—detecting all address\_reuse data races (2/2~RBs) versus its false negatives—through targeted path analysis. Crucially, our approach scaled to industrial projects: in \textit{Openethereum} (128k~LoC), we detected all 27 deadlocks with zero false positives and 74.4\% higher precision than LockBud (FP=3), resolving the tension between structural soundness and dynamic completeness that limits existing tools.

\begin{table*}[h]
\centering
\caption{Result on detection of deadlocks.}
\label{tab:improved_benchmark_table}
\begin{tabular}{|c|c|c|c|c|c|c|c|c|c|c|}
\hline
\multicolumn{2}{|c|}{\multirow{2}{*}{\textbf{Benchmark}}} & \multirow{2}{*}{\textbf{LoC}} & \multirow{2}{*}{\textbf{Lock}} & \multirow{2}{*}{\textbf{RB}} & \multicolumn{3}{c|}{\textbf{\#Deadlock}} & \multicolumn{3}{c|}{\textbf{Time (s)}} \\ \cline{6-11}  \multicolumn{2}{|c|}{}  & &   &   & \textbf{LockBud} & \textbf{Miri} & \textbf{Our} & \textbf{LockBud} & \textbf{Miri} & \textbf{Our} \\ \hline
\multirow{8}{*}{\rotatebox{90}{\textbf{L-Bench}}} 
 & conflict          & 75    & 3   & 1   & 1            & 3(FP:3)  & 1        & 0.05    & 15.89   & 4.9    \\ \cline {2-11}
 & conflict-inter    & 52    & 2   & 1   & 1            & 0        & 1        & 0.23    & 0.31    & 0.31   \\ \cline {2-11} 
 & intra             & 57    & 4   & 4   & 6(FP:2)      & 1(FN:3)  & 4        & 2.28    & 1.26    & 1.10   \\ \cline {2-11}
 & inter             & 178   & 6   & 12  & 14(FP:3,FN:1)& 1(FN:11) & 12       & 1.13    & 0.97    & 2.3    \\ \cline {2-11}
 & lock-closure      & 40    & 4   & 1   & 2(FP:1)      & 0(FN:1)  & 1        & 0.24    & 0.35    & 0.26   \\ \cline {2-11}
 & condvar-struct    & 273   & 8   & 4   & 1(FN:3)      & 2(FN:2)  & 4        & 2.01    & 1.17    & 1.77   \\ \cline {2-11}
 & static-ref        & 13    & 1   & 1   & 1            & 1        & 1        & 0.05    & 0.79    & 0.21    \\ \hline

\multirow{4}{*}{\rotatebox{90}{\textbf{M-Bench}}} 
 & singlethread  & 45   & 2   & 0   & 6(FP:6) & 0    & 0      & 0.18    & 1.8     & 0.08   \\ \cline {2-11}
 & mutex-leak         & 8    & 1   & 1   & 0(FN:1) & 1    & 1      & 0.10    & 0.33    & 0.16   \\ \cline {2-11} 
 & sync                & 266  & 7   & 1   & 1(FP:1)   & 0(FN)& 1      & 0.08    & 1.27    & 2.53   \\ \cline {2-11}
 & nopreempt     & 87   & 5   & 3   & 5(FP:2) & 3    & 3 & 0.26    & 0.31    & 0.28   \\ \hline

\multirow{4}{*}{\rotatebox{90}{\textbf{P-Bench}}} 
 & Rayon-demo       & 24k    & 72    & 2   & 0(FN:2)        & \ding{55}  & 2     & 4.18    & -   & 10.15    \\ \cline {2-11}
 & RCore            & 22k    & 263   & 8   & 8        & \ding{55}  & 8     & 7.82    & -   & 15.12    \\ \cline {2-11} 
 & Tikv-client      & 29k  & 4     & 0   & 0        & \ding{55}  & 0     & 34.12   & -   & 53.50    \\ \cline {2-11}
 & Openethereum     & 128k   &441    & 27  & 30(FP:3) & \ding{55}  & 27    & 45.32   & -   & 79.12    \\ \hline
\end{tabular}
\end{table*}


\begin{table*}[h]
\centering
\caption{Result on detection of data races.}
\label{tab:data_race_detection}
\begin{tabular}{|c|c|c|c|c|c|c|c|c|c|c|}
\hline
\multicolumn{2}{|c|}{\multirow{2}{*}{\textbf{Benchmark}}} & \multirow{2}{*}{\textbf{LoC}} & \multirow{2}{*}{\textbf{RB}} & \multirow{2}{*}{\textbf{UnsafeOp}} & \multicolumn{3}{c|}{\textbf{\#Race Condition}} & \multicolumn{3}{c|}{\textbf{Time (s)}} \\ \cline{6-11}  \multicolumn{2}{|c|}{}  & &   &   & \textbf{LockBud} & \textbf{Miri} & \textbf{Our} & \textbf{LockBud} & \textbf{Miri} & \textbf{Our} \\ \hline
\multirow{6}{*}{\rotatebox{90}{\textbf{M-Bench}}} 
 & datarace         & 226   & 1    & 34   & \ding{55}   & 1  & 1    & -  & 0.36    & 0.08   \\ \cline {2-11}
 & address-reuse     & 61    & 2    & 5    & \ding{55}   & 1(FN:1)  & 2    & -     & 16.02    & 0.36   \\ \cline {2-11} 
 & disable-datarace  & 31    & 1    & 5    & \ding{55}   & 1      & 1    & -     & 0.34    & 0.22   \\ \cline {2-11}
 & static-use         & 30   & 1    & 7    & \ding{55}   & 1      & 1    & -     & 0.31    & 0.49   \\ \cline {2-11}
  & arc-asptr         & 46   & 1    & 3    & \ding{55}   & 1      & 1    & -     & 149.41    & 0.49   \\ \cline {2-11}
 & simple             & 75    & 0    & 0    & \ding{55}   & 0      & 0    & -     & 0.08    & 0.22   \\ \hline 
\end{tabular}
\end{table*}

\begin{table*}[h]
\centering
\caption{Result on detection of atomicity violation.}
\label{tab:atomicity_violation}
\begin{tabular}{|c|c|c|c|c|c|c|c|c|c|c| }
\hline
\multicolumn{2}{|c| }{\multirow{2}{*}{\textbf{Benchmark}}} & \multirow{2}{*}{\textbf{LoC}} & \multirow{2}{*}{\textbf{Atomic}} & \multirow{2}{*}{\textbf{RB}} & \multicolumn{3}{c|}{\textbf{\#Race Condition}} & \multicolumn{3}{c|}{\textbf{Time (s)}} \\ \cline{6-11}  \multicolumn{2}{|c|}{}  & &   &   & \textbf{Lockbud} & \textbf{Miri} & \textbf{Our} & \textbf{LockBud} & \textbf{Miri} & \textbf{Our} \\ \hline
\multirow{5}{*}{\rotatebox{90}{\textbf{L/P-Bench}}} 
 & atomic-assert      & 226   & 8    & 1   & 0(FN:1)    & 0(FN:1) &  1  & 0.47    & 15.73   & 0.45   \\ \cline {2-11}
 & atomic-violation   & 72    & 5    & 4    & 4    & 0(FN:4)      &  4      & 0.18    & 17.03   & 0.08   \\ \cline {2-11}
 & atomic-closure         & 115   & 8    & 1    & 0(FN:1)    & 0(FN:1)     &  1      & 0.13    & 19.55   & 0.89   \\ \cline {2-11}
 & atomic-SeqCst          & 72    & 5    & 0    & 4(FP:4)    & 0     &  0      & 0.34    & 16.33   & 0.12   \\  \cline {2-11}
 & atomic-Ptr         & 294    & 7    & 2    & 0(FP:2)    & 2     &  0(FN:2)      & 2.42   & 1.64   & 2.85   \\  \hline
 
\end{tabular}
\end{table*}

\subsection{Research Questions}
While Petri nets offer formal foundations for concurrency modeling, their application to real-world Rust programs faces three ​fundamental challenges. To address these challenges and evaluate our integrated approach, we formulate the following research questions:To address the limitations of Petri nets in concurrency bug detection and validate our integrated approach, we formalize three research questions:  

RQ1: How can Rust's ownership and lifetime semantics be formally encoded into Petri net token flows while preserving bisimulation equivalence with the original program's behavior?

RQ2: What inherent limitations of Petri nets in detecting local deadlocks and atomicity violations can be overcome by integrating structural analysis with state reachability exploration?

RQ3: To what extent does our compression methods reduce the reachable state space without compromising detection completeness for deadlocks and data races?

\subsection{RQ1: Correctness of Ownership-Semantic Transformation}
Since the semantic equivalence of Rust programs and Petri nets within a subset of simplified grammars has been demonstrated in section \ref{transformation}, we can build on this conclusion to discuss the notions of soundness and completeness of transformations and subsequent analyses.

\begin{definition}[Rust Ownership in PPN]
\label{def:ownership-mapping}
The ownership system of Rust is mapped to PPN token semantics as follows:
\begin{itemize}
  \item Ownership: $M(p_r) = 1 \iff$ unique ownership of resource $r$.
  \item Borrowing: $\exists t \in Enabled(M(p_{r})),p_r \in ^{\bullet}t$ borrows of $r$.
  \item Lifetime: $lifetime(r) = [t_{start}, t_{end}] \iff M(p_r)\neq p_r.capticity$.
\end{itemize}
Transition preconditions enforce Rust's ownership rules: (1) $^{\bullet}t_{move}= \{p_x, p_r\}$;(2) $^{\bullet}t_{borrow} = \{p_x, p_{r}\}$ and $t_{borrow}^{\bullet} = \{p_x, p_{r}\}$.
\end{definition}

\begin{theorem}[Soundness of Concurrency Analysis]
\label{thm:concurrency-soundness}
The transformation from Rust programs to Petri nets preserves concurrency semantics, ensuring no false positives in detecting deadlocks, atomicity violations, and data races. Formally, for any program $Prog$ in our defined subset and its PPN:
\[
\lnot \exists \sigma \in \mathsf{Reach}(PPN): \sigma \vDash \varphi \implies \lnot \exists \tau \in \mathsf{Traces}(Prog): \tau \vDash \varphi
\]
where $\varphi \in \{\mathsf{deadlock}, \mathsf{atomicity\_violation}, \mathsf{data\_race}\}$. This guarantee stems from three pillars: (1) Control flow isomorphism ensures basic blocks map bijectively to places ($bb_f \leftrightarrow p_b$) and control operations to transitions ($\delta_f \leftrightarrow t_c$), preserving execution paths; (2) Resource binding consistency models synchronization primitives through dedicated places ($p_m$ for mutexes) and transitions ($t_{\mathtt{lock}}, t_{\mathtt{unlock}}$) with token semantics mirroring lock states; (3) Interleaving surjectivity encodes all possible thread schedules via concurrent transition firings in $T_{\mathtt{conc}}$, explicitly representing non-deterministic execution paths.
\end{theorem}

\begin{proof}
Soundness is established through operational correspondence: For deadlock freedom, if no siphon-trap imbalance ($\bullet S \subsetneq S\bullet$) exists in PPN, the program's wait-for graph contains no cycles. For data race absence, conflicting access pairs $(t_w, t_r)$ in $T_{\mathit{unsafe}}$ are always separated by synchronization transitions ($t_{\mathit{sync}}$) when required by program logic. Atomicity violations are prevented because atomic regions $\mathtt{Atomic}(ord)$ map to atomic transition sequences $[t_{\mathit{begin}}, \ldots, t_{\mathit{end}}]$ where intermediate firings are disabled by guard conditions. The proof leverages the control flow isomorphism which guarantees that for every program execution path $\tau$, there exists a firing sequence $\sigma$ in PPN with identical synchronization events and memory access ordering.
\end{proof}

\begin{theorem}[Completeness of Concurrency Bug Detection]
\label{thm:concurrency-completeness}
The PPN model detects all concurrency bugs present in the original program:
\[
\exists \tau \in \mathsf{Traces}(Prog): \tau \vDash \varphi \implies \exists \sigma \in \mathsf{Reach}(PPN): \sigma \vDash \psi(\varphi)
\]
with defect mapping $\psi$: 
\begin{align*}
\psi(\mathsf{deadlock}) &\triangleq \exists S \subseteq Prog: \bullet S \subsetneq S\bullet \\
\psi(\mathsf{atomicity\_violation}) &\triangleq \exists t_{\mathit{load}} \in [t_{\mathit{begin}}, t_{\mathit{end}}]: \mathsf{fire}(t_{\mathit{load}}) \\
\psi(\mathsf{data\_race}) &\triangleq \exists t_w, t_r \in T_{\mathit{atom}}: \mathsf{concurrent}(t_w, t_r) \land \neg \mathsf{unsafe}(p_r)
\end{align*}
Completeness holds for bounded thread creation and finite recursion depth.
\end{theorem}

\begin{proof}
Completeness follows from state space coverage and defect manifestation: (1) Deadlocks manifest as token starvation in siphons where resource places $p_m$ are held while waiting places $p_{\mathtt{wait}}$ have tokens but cannot fire due to missing resources; (2) Data races occur when conflicting atomic transitions ($t_w$ for write, $t_r$ for read) fire concurrently without mutual exclusion tokens in $p_m$; (3) Atomicity violations appear when intermediate transitions $t_{\mathit{load}}$ fire between $t_{\mathit{begin}}$ and $t_{\mathit{end}}$ of an atomic region, detectable through forbidden state patterns. The reachability graph covers all thread interleavings by construction, as each spawn operation ($\mathtt{thread::spawn}$) generates new tokens in thread-specific places, enabling concurrent firing of transitions in different thread subnets. Partial-order reduction preserves bug traces while mitigating state explosion.
\end{proof}


\subsection{RQ2: Effectiveness on Detecting Concurrency Bugs}
The results in Table~\ref{tab:improved_benchmark_table} demonstrate the performance of our tool compared to LockBud and Miri across various benchmarks, including L-Bench, M-Bench, and P-Bench. Miri, as a Rust interpreter, exhibits robust runtime concurrency detection capabilities, supporting asynchronous operations and multiple concurrency primitives. However, it is limited to executable crates, making it incapable of directly analyzing library crates without executable test cases. Furthermore, its runtime reports are often challenging to interpret, and its reliance on incomplete models of certain concurrency scenarios can lead to false negatives, as shown in benchmarks such as condvar-struct and inter. 

\begin{figure*}[!htb]
    \centering
    \subfigure[Example of a false positive.]{\includegraphics[width=0.42\textwidth]{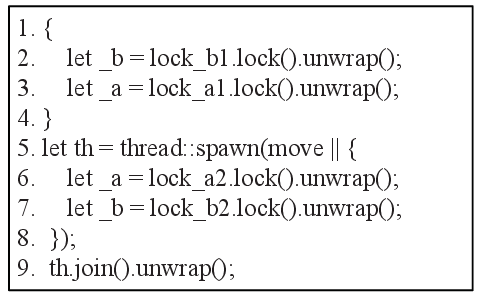}\label{fig:lockbud-fp} }\hspace{1cm}
    \subfigure[Example of a false negative.]{\includegraphics[width=0.35\textwidth]{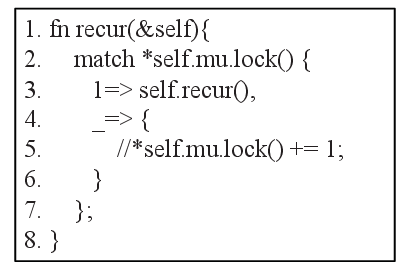}\label{fig:lockbud-fn}}
    
    \caption{Examples of false positives and false negatives in Lockbud.} 
    \label{fig:lockbud_examples}
\end{figure*}

LockBud is a lightweight tool based on lock graph analysis, offering efficiency in detecting locking-related issues. Nevertheless, it suffers from two primary sources of false positives: first, its conservative assumption that multiple read operations on a RwLock could block subsequent write locks; second, its inability to accurately model thread states, leading to lifecycle analysis errors for locks with different aliases. These shortcomings are evident in its results for benchmarks like intra and conflict-inter. Additionally, LockBud struggles with condvar-based synchronization mechanisms, as it cannot definitively determine whether a notify signal has already occurred prior to a thread's wait, resulting in both false positives and false negatives. Fig. \ref{fig:lockbud_examples} illustrates examples of both false positives and false negatives reported by Lockbud. The first type of false positives in Lockbud is due to a lack of thread modeling. In Fig. \ref{fig:lockbud-fp}, the thread creation function is called in line 5. Due to the lifetime constraint, the code from lines 1 to 4 has been executed and the lock held by it has been released, so there is no competition for lock resources after thread creation, there is no deadlock. Moving lines 5 to 9 of the code before line 1 will not result in a deadlock, but Lockbud will still produce a false positive. The false negative occurs because Lockbud lacks proper modeling of recursive calls, mistakenly assuming that locks in recursive calls have already been released during its lifetime analysis.

In comparison, our Petri net-based approach leverages state reachability analysis, allowing for precise modeling of concurrency scenarios. By specifying function entry points and systematically exploring potential deadlock states, our tool achieves comprehensive coverage while maintaining accuracy, as evidenced by its performance in detecting all true deadlocks in benchmarks like inter and lock-closure. Furthermore, our tool produces precise reports with no false positives across the evaluated benchmarks.

The results in Table~\ref{tab:data_race_detection} and Table~\ref{tab:atomicity_violation} demonstrate the detection of data races and atomicity violations, as these two categories share common characteristics. Atomicity violations are typically caused by relaxed memory orderings, leading to concurrent load or store operations across threads. These issues can also be considered as a special case of race conditions. As such, we evaluate both types of bugs together. LockBud does not support the detection of data race bugs. For atomicity violation detection, it only focuses on the sequential usage patterns of atomic APIs within a single thread. Consequently, it fails to detect atomicity violations in multithreaded scenarios. Moreover, even when changes are made to memory ordering or additional locks are added, LockBud still incorrectly reports the presence of atomicity violations, reflecting limitations in its modeling of synchronization and memory order semantics. Miri relies on runtime analysis. While effective in detecting race conditions and atomicity violations, it is prone to termination errors if unwrap operations are not preemptively handled. Additionally, due to its reliance on executed paths, some concurrency issues may be missed, particularly if the problematic execution path is not covered during testing. Our Petri net-based framework exhibits significant advantages in both data race and atomicity violation detection. By explicitly modeling transitions for unsafe data operations, our framework not only detects traditional race conditions but is also theoretically capable of identifying more complex concurrency bugs such as use-after-free (UAF) and null pointer dereferences. However, in this study, we focus specifically on race conditions where at least one operation on the same unsafe data is a write. This approach extends naturally to the detection of atomicity violations by capturing relaxed memory ordering operations and store that occur concurrently without appropriate synchronization mechanisms. Overall, the results highlight the flexibility and precision of our approach in detecting both data races and atomicity violations, demonstrating its superiority over LockBud and Miri in addressing multithreaded and relaxed-memory scenarios. However, as the detection of atomicity violations is inherently challenging in multithreaded environments, further refinement of the Petri net's modeling capabilities could further improve its accuracy and applicability.

\begin{figure*}[!htb]
    \centering
    \subfigure[Example in AtomVChecker.]{\includegraphics[width=0.4\textwidth]{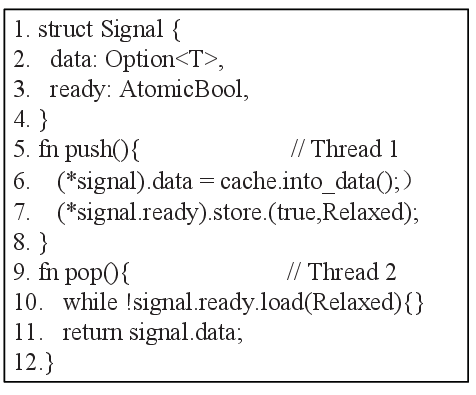}\label{fig:atomvchecker}}\hspace{1cm}
    \subfigure[Example in PPN.]{\includegraphics[width=0.4\textwidth]{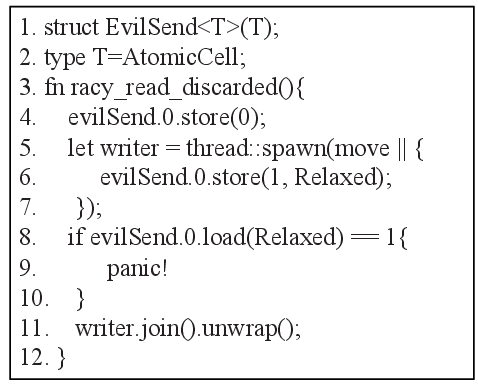}\label{fig:atomppn}}
    
    \caption{Examples of AtomVChecker and PPN detecting atomicity violations.} 
    \label{fig:atomic_example}
\end{figure*}
For example, another existing atomicity violation detection tool, AtomVChecker~\cite{wang2024understanding}, identifies several patterns of atomicity violations in Rust, as shown in Fig. \ref{fig:atomvchecker}. These include operations like push and pop performed in different threads, where relaxed memory ordering may cause pop to return an uninitialized variable. The PPN approach discussed in this paper primarily detects race conditions caused by multiple store operations, leading to undefined behavior during a load. AtomVChecker cannot report errors in Fig. \ref{fig:atomppn}, while PPN cannot detect the atomicity violations in Fig. \ref{fig:atomvchecker}. The two approaches can complement each other. In future work, we will use advanced nets to represent the dependencies between data.

\subsection{RQ3: Efficiency of State Compression in Bug Detection}
The proposed reduction methods demonstrate exceptional efficiency in bug detection by fundamentally addressing state explosion through three complementary techniques: (1) linear sequence compression eliminates intermediate states while preserving firing sequences, (2) sensitive call chain pruning removes concurrency-irrelevant paths while protecting shared nodes, and (3) localized loop abstraction breaks cycles without resource access—collectively achieving average structural reductions of 34.88\% for places, 37.21\% for transitions, and 33.76\% for flow relations while enabling superlinear state space compression (e.g., Rayon-demo's 94.5\% reduction from 1259 to 69 states), where the remarkably small post-reduction state spaces ($States^{\prime}$) directly result from deadlock-induced state space truncation during coverability graph generation, as deadlock states become terminal nodes that prevent further state expansion; crucially, the methods' formal preservation of safety and liveness properties ensures all concurrency bugs—including the atomic violations detected in the 123-state reduced model (from 127 original states) and deadlocks in condvar-lock (116→32 states)—remain verifiable while reducing verification complexity from exponential to near-linear, with maximal processing time of just 4.8ms for industrial-scale models (2316 elements), thereby overcoming the fundamental state explosion barrier that traditionally limits Petri net verification.

\begin{table}[htbp]
\centering
\caption{Effectiveness Evaluation of Petri Net Reduction Methods}
\label{tab:reduction_results}
\begin{tabular}{|c|c|c|c|c|c|c|c| }
\hline
\multirow{2}{*}{Benchmarks} & 
\multicolumn{2}{|c| }{Original Model} & 
\multicolumn{5}{|c| }{Reduction Effectiveness (\%)} \\
\cline{2-3} \cline{4-8}
& {|P|+|T|+|F| } & {States} & {Places} & 
{Transitions} & {Flow} & {Time(ms)}  & {$States^{\prime}$}\\
\hline
intra & 313 & 55 & 20.00 & 21.43 & 17.86 & 1.5  & 48  \\
inter & 168 & 40 & 38.10 & 44.74 & 38.64 & 0.6  & 12 \\
condvar-lock & 210 & 116 & 41.51 & 44.68 & 40.91 & 1.5 & 32  \\
atomic-violation & 555 & 127 & 24.03 & 27.56 & 23.75 & 1.3 & 123 \\
Rayon-demo & 2316 & 1259 & 48.29 & 50.80 & 47.66 & 4.8 & 69  \\
\hline
\textbf{Average Reduction} & - & - & - & 34.88 & 37.21 & 33.76 & -  \\
\hline
\end{tabular}

\vspace{0.5em}
\footnotesize
\begin{itemize}
  \item All values represent percentage reduction: $\frac{\text{Original} - \text{Reduced}}{\text{Original}} \times 100$
  \item Combined reduction applies all three methods sequentially
\end{itemize}
\end{table}


\section{Conclusion}
\label{conclusion}
In this work, we proposed a formal framework for detecting concurrency bugs in Rust programs using Petri nets. By defining a precise program model and converting it into a PPN representation, we enabled the systematic analysis of deadlocks, data races, and atomicity violations. The formal definitions of these bug patterns, coupled with RG traversal and tailored detection algorithms, ensure both the soundness and completeness of our approach. To improve the scalability of the framework, we introduced a Petri net simplification algorithm that reduces the size and complexity of the net while preserving its execution semantics. This optimization enhances the efficiency of the bug detection process without sacrificing accuracy. Through the rigorous construction of resource places, transitions, and state-based analysis, our method captures the essential synchronization and unsafe operation semantics of Rust programs. By leveraging the expressiveness of Petri nets, this framework provides a unified and formalized approach to identifying key concurrency bugs, ensuring correctness and reliability in the detection process. Future work will focus on extending this framework to handle more advanced concurrency patterns, improving the performance of state exploration algorithms, and applying the methodology to real-world Rust programs to validate its effectiveness and scalability.

\bibliographystyle{ACM-Reference-Format}
\bibliography{sample-base}

\end{document}